# Quartz Plate Calorimetry for CMS HE Upgrade

Yasar Onel* and David Winn**

for the US-HE Upgrade Group[1]

*University of Iowa, Iowa City, IA, USA
**Fairfield University, Fairfield, CT, USA

**Abstract**

Analysis of the CMS data and the simulation prediction based on these results indicate that the performance of the current scintillators in the CMS Hadron Endcap Calorimeter (HE) tiles will degrade dramatically in the High Luminosity LHC (HL-LHC) era. In order to continue the physics program in this region, the HE tiles will need to be replaced. The new tiles should have comparable/improved performance, be radiation hard, reliable and robust.

## 1. Introduction

Hadronic Endcap (HE) calorimeters of the CMS experiment cover the pseudorapidity range of $1.4 < \eta < 3$ on both sides of the CMS detector, contributing to jet and missing transverse energy resolutions. It was observed in the first round of LHC running between 2008-2012 that the scintillator tiles used in the CMS Hadronic Endcap calorimeter will lose their efficiency and will not be able to survive the High-Luminosity LHC (HL-LHC) running conditions.

The CMS collaboration plans to substitute Quartz plates for the scintillator tiles of the original design. Various tests have proved Quartz to be radiation hard, but the light produced by Quartz comes from Cerenkov process, which yields drastically fewer photons than scintillation. To increase the light collection efficiency, we propose to treat the Quartz plates with radiation hard wavelength shifters, p-terphenyl or 4% gallium doped zinc oxide. The test beam studies revealed a substantial light collection increase on pTp or ZnO:Ga deposited Quartz plates. We con-

---

[1] US-HE Upgrade Group: Iowa, Baylor, Fairfield, Fermilab, FIU, Maryland, Mississippi, Notre Dame

Extended US Group for HCAL Upgrades: Boston, Minnesota, Princeton, Virginia

International Group: Rio-CBPF,Brazil; Rio-UERJ,Brazil; Sao Paulo-Unicamp,Brazil; INFN Trieste, Italy; Bogazici U. Istanbul,Turkey; Cukurova U, Adana, Turkey; ITU, Istanbul, Turkey; METU,Ankara, Turkey structed a 20 layer calorimeter prototype with pTp coated plates, and tested the hadronic and the electromagnetic capabilities at the CERN H2 area. Here we describe the outcome of various tests in different configurations using quartz plates as active elements in calorimetry.

Details of the radiation damage tests of the quartz plates and the fibers and the beam tests can be found in [1-7].

## 2. Quartz Plates with WLS Fibers

This technology utilizes quartz plates with Wavelength Shifting (WLS) Fibers running in grooves of different geometries read out with photodetectors as the active medium. The selected fiber geometry, which is optimized with simulations, is the so-called "bar-shaped" geometry where the fibers run along the parallel grooves that are 2 cm apart, bundled on one side and read out with photodetectors. A calorimeter prototype was built and tested at Fermilab test beam. The calorimetric performance is comparable with the current HE. Dedicated R&D on the radiation hard WLS fibers points towards pTp cladded quartz fibers and pTp/anthracene filled quartz capillaries.

Figure 1a shows the so-called bar-shaped fiber geometry in the quartz plate and Fig. 1b shows the results of the first tests using the anthracene filled capillaries. In these tests, an array of seven capillaries was exposed to the electron shower at its maximum and was read out with two PMTs at both ends. The peak corresponds to 7 photoelectrons.

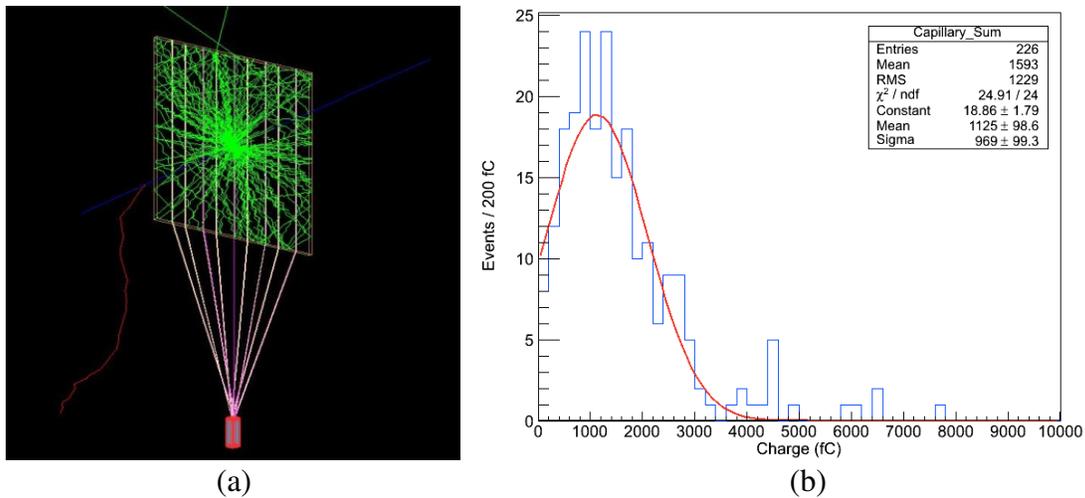

(a)          (b)

**Figure 1.** The quartz plate with the bar-shaped fiber geometry (a) and the signal from the first tests of anthracene filled capillaries in an electron shower (b).

## 3. Quartz Plates with Inorganic Scintillators

In this approach, the quartz plate is coated with an inorganic scintillator (pTp, ZnO:Ga, oTp, mTp or pQp) and read out with a photodetector directly coupled to the side of the plate. All scintillators yielded the same performance with pTp being the easiest to apply. A calorimeter prototype with pTp coated quartz plates was built and tested at CERN test beam. The calorimetric performance is suitable for the HE operations and the agreement with simulation is satisfactory.

Figure 2 shows the hadronic response and the energy resolution of the prototype.

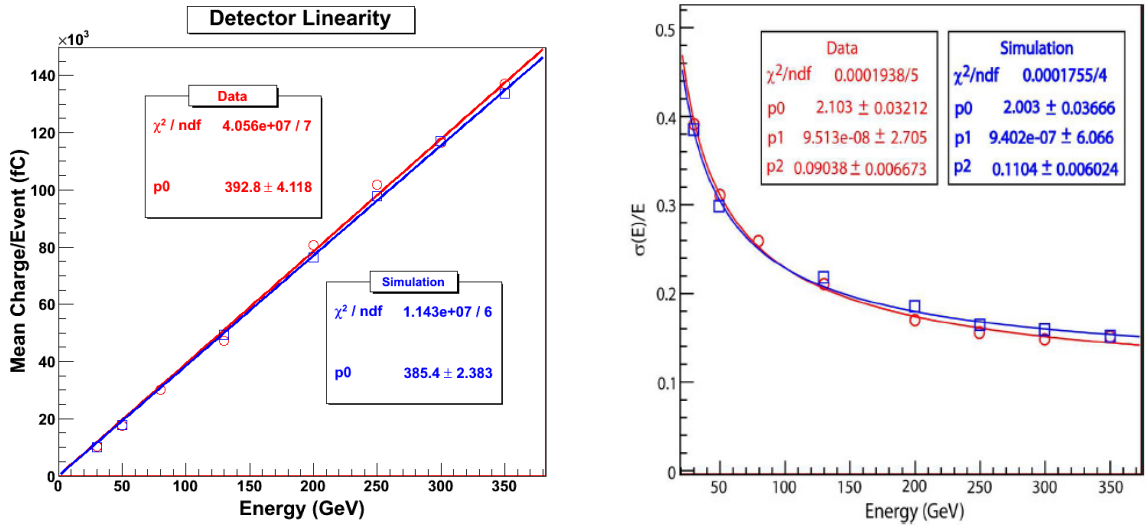

**Figure 2.** Hadronic response (left) and the energy resolution (right) of the calorimeter prototype built with pTp coated quartz plates.

The quartz plate technology is a viable option in the case that the upgrade decision about the HE is the replacement of the scintillator tiles with radiation hard, equally well performing alternatives.

## 4. Summary

This R&D is for developing radiation-hard active elements for hadron calorimeters. Here we choose the quartz plates as the base technology and investigate novel techniques for the enhancement of the optical signal and the readout. The R&D also enhances the understanding of inorganic scintillators and radiation-hard WLS fibers.

## References

**1.** Dumanoglu *et al.* "Radiation-hardness studies of high OH content quartz fibers irradiated with 500 MeV electrons" Nucl. Instr. Meth. A 490 (2002) 444-455

**2.** Cankocak *et al.* "Radiation-hardness measurements of high OH content quartz fibres irradiated with 24 GeV protons up to 1.25 Grad" Nucl. Instr. and Meth. A 585 (2008) 20–27

**3.** U. Akgun *et al.* "Radiation Damage in Quartz Fibers Exposed to Energetic Neutrons" "Radiation Damage in Quartz Fibers Exposed to Energetic Neutrons", submitted to IEEE Transactions on Nuclear Science.

**4.** F. Duru *et al.* "CMS Hadronic EndCap Calorimeter Upgrade Studies for SLHC - Cerenkov Light Collection from Quartz Plates", IEEE Transactions on Nuclear Science, Vol 55, Issue 2, 734-740, 2008.

**5.** U. Akgun *et al.*, "Quartz Plate Calorimeter as SLHC Upgrade to CMS Hadronic Endcap Calorimeters", XIII International Conference on Calorimetry in High Energy Physics, CALOR 2008, Pavio, Italy, May 2008, J.Phys.Conf.Ser.160:012015, 2009

**6.** U. Akgun *et al.* "CMS Hadronic Calorimeter Upgrade Studies - P-Terphenyl Deposited Quartz Plate Calorimeter Prototype ", APS 2009, Denver, CO, USA, May 2009.

**7.** B. Bilki *et al.* "CMS Hadron Endcap Calorimeter Upgrade Studies For SuperLHC", CALOR 2010, Beijing, China.